\documentclass{article}
\usepackage{authblk}
\usepackage[toc,page]{appendix}
\usepackage[left=1.5cm,right=1.5cm,top=2cm,bottom=2cm]{geometry}
\usepackage{amsmath,amsbsy,amsfonts,amssymb}
\usepackage[utf8x]{inputenc}
\usepackage{graphicx}
\usepackage{wasysym}
\usepackage{multicol}
\usepackage{fancyhdr}
\usepackage{hyperref}
\usepackage{color}
\usepackage{float}
\usepackage{enumitem}
\usepackage{subfloat}

\author[1]{Irving  Rond\'on \footnote{corresponding author: irondon@kias.re.kr}  }
\affil[1]{School of Computational Sciences, Center for  In Silico Protein Science,}
\affil[1]{Korea Institute for Advanced Study, 
	85 Hoegi-ro, Seoul 0245, Republic of Korea}

\author[2]{Oscar  Sotolongo-Costa}
\affil[2]{Universidad Aut\'onoma del Estado de Morelos,  C.P. 62209,
Cuernavaca, M\'exico}
\author[3]{Jorge  A. Gonz\'alez}
\affil[3]{Department of Physics, Florida International University, Miami, Florida 33199, USA} 

\author[1]{Jooyoung Lee}

\title{A generalized $q$ growth model based on nonadditive entropy}

\begin{document}

\maketitle

\begin{abstract}
	We present a general growth model based on nonextensive statistical physics. We show
	that the most common unidimensional growth laws such as power law, exponential,
	logistic, Richards, Von Bertalanffy, Gompertz can be obtained. This model belongs to
	a particular case reported in (Physica A 369, 645 (2006)). The new evolution equation
	resembles the “universality” revealed by West for ontogenetic growth (Nature 413, 628
	(2001)). We show that for early times the model follows a power law growth as  $ N(t) \approx t ^ D $, where the exponent $D \equiv  \frac{1}{1-q}$ classifies different types of  growth. Several examples are given and discussed.

\end{abstract}

\section{Introduction}
The application of the concept of entropy in mathematical modeling has recently had a  significant impact in different scenarios such as solid state physics \cite{Miracle}, optics \cite{Alfonso}, medicine \cite{ABarrea} and many other fields \cite{James1,James2,Oden,Lima,Tsoularis,Harold}.
\\
\\
Nowadays, several research groups have been working in proposed generalized 
growth models able to describe several features as the exponential, Gompertz, logistic, power law, just to mention some of them.
\\
\noindent
Recently, mathematical modeling and computational simulations have been the most effective and useful tools that help universal global efforts to understand coronavirus disease (COVID-19). Currently, the world official data for the time evolution of active cases for this pandemic around the world are available online  \cite{site1} \cite{site2} \cite{site3}.
\\
\\
In the  context of non-extensive entropy, C. Tsallis has proposed a  new $q$ statistic distribution to describe the time evolution of the number of patients with COVID-19 in order to predict per day the  height of the peak per country \cite{Constantino} and  references therein. \\ 
Therefore, the model presented here has a direct connection to model COVID-19  to predict the evolution growth, but this problem will be addressed for the authors somewhere due space, complexity and details.
\\
\\
In this letter, we focus on the work proposed by
West  \cite{West}, where the authors derived a general ontogenetic quantitative model based on fundamental principles, and obtained  a single universal curve that describes the growth of many diverse species using experimental data.
\\
\\
In Ref. \cite{Delsanto} an analytic approach to obtain the well known Gompertz, West, and logistic models were introduced. This idea was successfully applied  in  tumor growth  \cite{Guiot}, modeling ecological systems \cite{Barberis},
and  vector growth universality \cite{Barberis2}.
\\
The use of nonextensive statistics for modeling cancer laws and treatment purposes was reported \cite{JGIRondon,Jbanda}. This approach has been applied 
for radiotherapy treatment  schedules \cite{Barrea}.
\\
Continuous and discrete growth models based in $q$ function formalism \cite{Martinez,Martinez1} have been proposed to understand the interaction and the evolution of complex systems \cite{Barberis3}, and the modeling problem of power exponent growth universality \cite{Luciano,Guiot1}.
\\
\noindent
Due to the complexity in natural phenomena, those models must be intrinsically nonlinear. 
\\
\\
Empirical hypotheses can be proposed, for example, cellular, structures, and cancer themselves possess fractal behavior \cite{Spillman}.
\\
\\
The latter issue has been studied to get a deeper understanding of a particular system. It has been shown to lead various observed growth laws such as exponential, logistic, Richards, Gompertz, which may possess fractal behavior \cite{Monbach}.
\\
\\
In this context, we  have used the non-extensive  entropy proposed by Tsallis \cite{Tsallis}. This theory  has been supported by many successful applications \cite{ABE,Tsallis2}, and it has been used to  explain several phenomena
possessing nonlinear behavior, long-range memory and  fractal interactions that are present in nature. 
\\
This theory opens the possibility to study growth problems from a different point of view
\\
\\
The main goal  in this contribution is to show that by using simple arguments of statistical mechanics a generalized  $q$  growth model encompassing exponential, logistic, Gompertz-like growth and power laws.\\ 
In this paper, we present a new growth model that resembles a universal ontogenic growth proposed by West but here it is directly related with non-extensive statistics.
\\
\\
The paper is organized as follows, in section
2   the non-extensive statistics is briefly presented, in section 3,  using Tsallis theory, a generalized growth model is proposed so that particular cases such a power law, exponential, logistic, Richards, Von Bertalanffy, and Gompertz laws. In section 4,  we derive an explicit expression  useful  in order to  obtain the fractal dimension for this model, finally some conclusions are given.
\section{Tsallis entropy}
Since 1988 Constantino Tsallis \cite{Tsallis} proposed a possible generalization of  Boltzmann-Gibbs (BG) statistics.
\\
Arguing that the canonical BG entropy definition is not universal and valid for systems where long-range interactions are not meaningful, Tsallis proposed a generalization of the BG functional trying to incorporate non-euclidean geometry effects and long-range interactions.\\
This generalized entropy, known as Tsallis entropy (denoted below as $S_q$), is dependent on a new parameter $q$, which keeps tracks of non BG effects present in the system.
\begin{equation}
\label{TsallisEntropy}
S_q = k \frac{1- \sum_{i=1}^{w} P_{i}^q }{q-1},
\end{equation}
where $\kappa$ is a proportionality factor, $w$ is the total number of possibilities of the system,
$ \sum_{i=1}^{w} P_{i} = 1$, $q\,\,\epsilon \, R$. \\
Expression (\ref{TsallisEntropy}) recovers the BG entropy,
$S_1 = - k \sum_{i=1}^{w} P_{i}\ln P_{i}$ in the limit $q\,\, \rightarrow 1$, while the parameter $q$ characterizes the degree of nonextensivity of the system.
\\
On the other hand this entropy follows the rule
\begin{equation}
\label{SumRule}
S_{q}(A+B)/k = [ S_{q}(A)/k] + [ S_{q}(B)/k]  + (1 - q)[ S_{q}(A)/k][ S_{q}(B)/ k],
\end{equation}
where $A$  and $B$ are two independent systems. In equation (\eqref{SumRule}) for $q=1$ the extensive  entropy summability is recovered. 
\\
This generalized statistic has led to a large amount of successful applications \cite{ABE}. 
\\
These include systems with long-range interactions, long-range microscopic memory, and systems with a fractal or multi-fractal structure.
\\
Details from this theory and its results can be reviewed \cite{Tsallis2} and references therein.
\section{Nonextensive entropy  can support universal growth law}
A general growth model can be derived using energy balance considerations \cite{West}. This idea was applied to study tumor growth \cite{Delsanto,Guiot}.\\
However,  the authors did not use statistical mechanics in their formulations.
\\
In this sense, using \cite{Calderon} the authors used  BG entropy to find the Gompertzian growth. \\
Following this approach the authors in \cite{JGIRondon} derived  a new growth model  using non-extensive entropy.
\\
In the context of tumor  or evolution growth, the idea was to define states of probability. 
\\
For active (reproducing) cells $P_1$ and for resting cells $P_2$, the nonextensive entropy given by the equation  \eqref{TsallisEntropy}  can be written as
\begin{equation}
\label{ecu1}
S=  \frac{k }{q-1} \left( 1- P_{1}^q - P_{2}^q \right).
\end{equation}
Considering them as  measurable observable, the probabilities can be normalized, using the following approximation  $P_1 \equiv N(t)/N_{\infty}$,  which stands for the ratio between a  population $N(t)$ and the total  size or volume (and the carrying capacity) expressed as $N_\infty$, considered as constant. $P_2 \equiv 1-P_1$ is the fraction of inactive cells. Assuming proportionality between population growth rate and entropy \cite{Calderon} but using Tsallis entropy the  following equation was presented  \cite{JGIRondon}.
\begin{equation}
\label{Eq4}
\frac{dN}{dt} = \frac{k N_{\infty} }{ q -1 } \left [ 1 -  \left (\frac{N}{N_{\infty}}\right)^q - \left( 1 -\frac{N}{N_{\infty}}\right)^q \right ].
\end{equation}
This dynamical equation \eqref{Eq4} predicts  different growth curves for particular $q$  values \cite{JGIRondon}.
\\
This assumption can be approximately valid since the entropy equation \eqref{TsallisEntropy} has been primarily applied to different physical problems, including different models and simulations, e.g.,
\href{http://www.cbpf.br/GrupPesq/StatisticalPhys/biblio.htm}{http://www.cbpf.br/GrupPesq/StatisticalPhys/biblio.htm}
  for a substantial bibliography, leading to good agreements with experimental data \cite{WebSite}.
\\
For instance, the idea and main contribution in this work  is to obtain a general "universal"  growth model as \cite{West}, but using non-extensive entropy \cite{JGIRondon}.
\\ 
Therefore, if we consider small active cells $N/N_\infty << 1$, it corresponds in our case to early states of growth, this is an essential condition that is pointed out and discussed in detail to developed, validate and calibrate a mathematical model in tumor growth \cite{Oden,Lima}. 
\\
Then, $P_2^q$ can be approximated through a Maclaurin series on  $N/N_{\infty}$, as  $P_2^q = 1 - q (N/N_{\infty})  + O((N/N_{\infty})^2)$ up to first order.
Under this approximation, we show in the following sections  that different growth behaviors can be obtained.
\\
This approximation is supported in  \cite{Monbach} when initial early growth is slow. In our case, it allows us to maintain only the first term of this expansion.
\\
After substituting the above  approximation into Eq. \eqref{Eq4} it is straightforward to  obtain  the  following evolution equation 
\begin{equation}
\label{newequation}
\frac{dN}{dt}=   \alpha_q \left(\frac{N}{M_{\infty}}\right)^q\left[1  -  \left(\frac{N}{M_{\infty}}\right)^{1-q}\right],
\end{equation}
where  $\alpha_q \equiv k \frac{q M_{\infty} }{1-q} $ can be considered  as  a kinetic constant dependent on $M_{\infty}\equiv q^{\frac{1}{q-1}} N_\infty$ which is related with  the carrying capacity. This equation constitute the main results of the present analysis.

In the next section, we illustrated how this parameter varies for different growth laws,  
It is essential to remark  that the new equation \eqref{newequation} resembles the universal growth law reported in \cite{West}, but here the possibilities of non-extensive interpretation for the first time.

\begin{figure}
	\centering
	\includegraphics[scale=0.85]{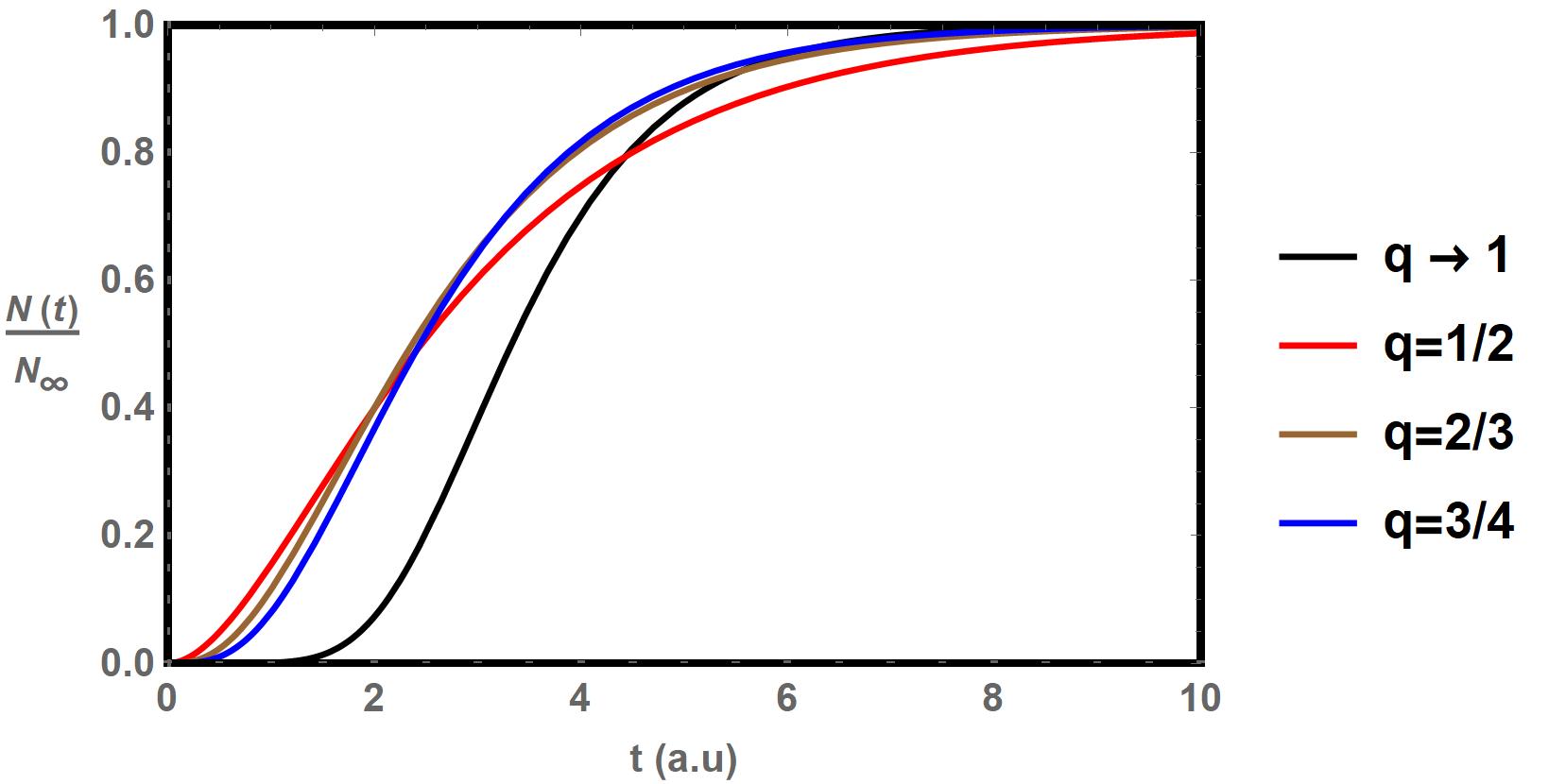}
	\caption{ Normalized numerical solution Eq. \eqref{qUniversal}  for different growth models.}
	\label{Fig:figura0}
\end{figure}

\subsection{Examples of classical growth models for different $q$ values}
\label{sectionqModel}
After substituting $q \rightarrow 1$ into \eqref{newequation}, we obtain the Gompertz-like  growth
reported in \cite{Bajzer,Norton}
\begin{align}
\label{Gompertz}
\frac{dN}{dt}=  kN- kN \ln \left(\frac{N}{N_\infty}\right)
\end{align}
A particular  power law growth can be obtained when  $q=1/2$. As an example   breast cancer is described in \cite{Hart}
\begin{align}
\label{Hart}
\frac{dN}{dt}= 2 k N_\infty^{1/2} N^{1/2}- kN
\end{align}
The Bertalanffy model \cite{Bertalanffy} can be obtained for $q=2/3$. 
\begin{align}
\label{Bertalanffy}
\frac{dN}{dt}= 3 k N_{\infty}^{1/3} N^{2/3}-2 k N
\end{align}
For the case  $q=3/4$ West model,  it  fits the data gathered from a wide range of living systems such as mammal, birds, mollusks and plants \cite{Dodds,West}, this behavior  has been recently reviewed and reported \cite{Fernando}.
\begin{align}
\label{West}
\frac{dN}{dt}= 4 k{N_\infty}^{1/4} N^{3/4}-3 k N
\end{align}
All of these growth laws are well known in the literature \cite{Farkas} and references therein. \\
In Figure~\ref{Fig:figura0}, we  show the numerical solution 
for different $q$ values. 

\begin{figure}
	\centering
	\includegraphics[scale=0.85]{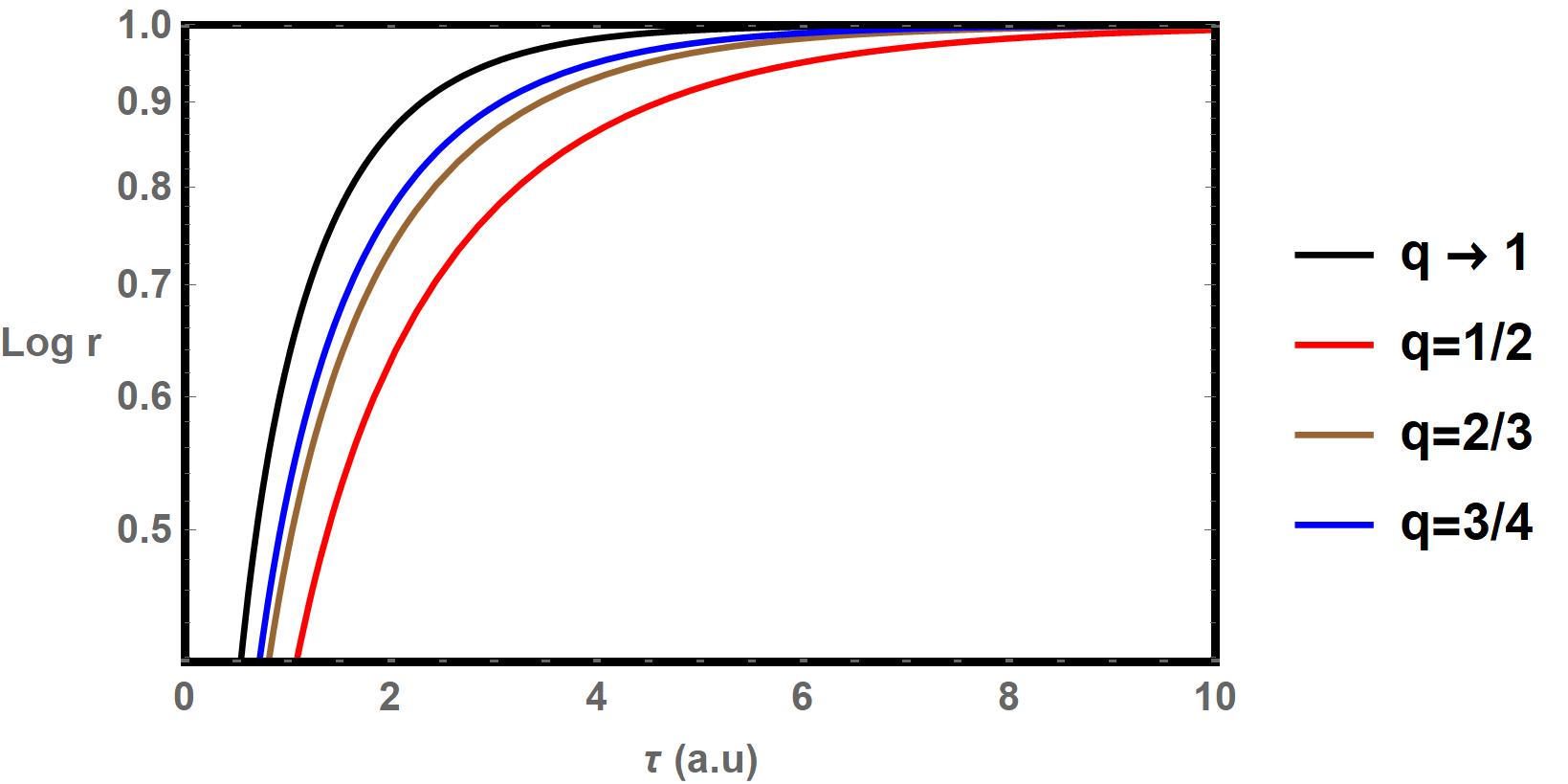}
	\caption{ Log Universal  parameterless $q$ curve Eq. \eqref{newequation}  for different growth models.}
	\label{Fig:figura1}
\end{figure}
\section{A classical generalized $q$ sigmoidal curve}
Through the analytical solution given by the $q$ growth model  \eqref{newequation}, with the initial condition  $N(t=0)=N_{0}$, it is straightforward to obtain a generalized type of West model \cite{West}. The solution can be written as 
\begin{equation}
\label{universal}
\left(\frac{N}{M_{\infty}}\right)^{1-q} =1-\left[1- \left(\frac{N_0}{M_{\infty}}\right)^{1-q}\right] e^{{-q k t}}.
\end{equation}
Let us  define  $r \equiv  \left( \frac{N}{M_{\infty}} \right)^{1-q} $ and  $\tau \equiv k t  -  \ln  \left( 1- r_0  \right)^{ \frac{1}{q}}  $, where $r_0 \equiv  \left( \frac{N_0}{M_{\infty}} \right)^{1-q} $, after substitution into Eq. \eqref{universal}  we obtain  a  growth equation
\begin{equation}
\label{qUniversal}
r = 1 - e^{-q\tau}
\end{equation}
A similar expression was derived as a single 
curve that describes the growth of many diverse species in \cite{West}, where the authors  showed that the same universal exponential curve fits the ontogenetic growth data on mammals, birds, fishes, and mollusks, providing masses and growth times for the different organisms with properly classical allometric scaling. \\ This kind of model was used to fit the growth of tumor spheroids in vitro and patient data, reported in \cite{Condat1}, it has  also used in  \cite{Renato}  to  study avascular tumor growth.
\\
\\
Eq. \eqref{qUniversal}  suggests a  connection between non-extensive statistics and the growth laws. 
The $q$ value can be related,
to describe different lines of tumor growth  \cite{Guiot,Guiot1}.  \\
A similar approach can be done using a nonlinear fitting, and it can be applied when the biological meaning of the $q$ parameter is unknown, but this analysis goes beyond this contribution. 
\\
In Figure~\ref{Fig:figura1}, we  show the behavior of Eq. \eqref{qUniversal} for different $q$ values. 
\begin{figure}
	\centering
	\includegraphics[scale=0.85]{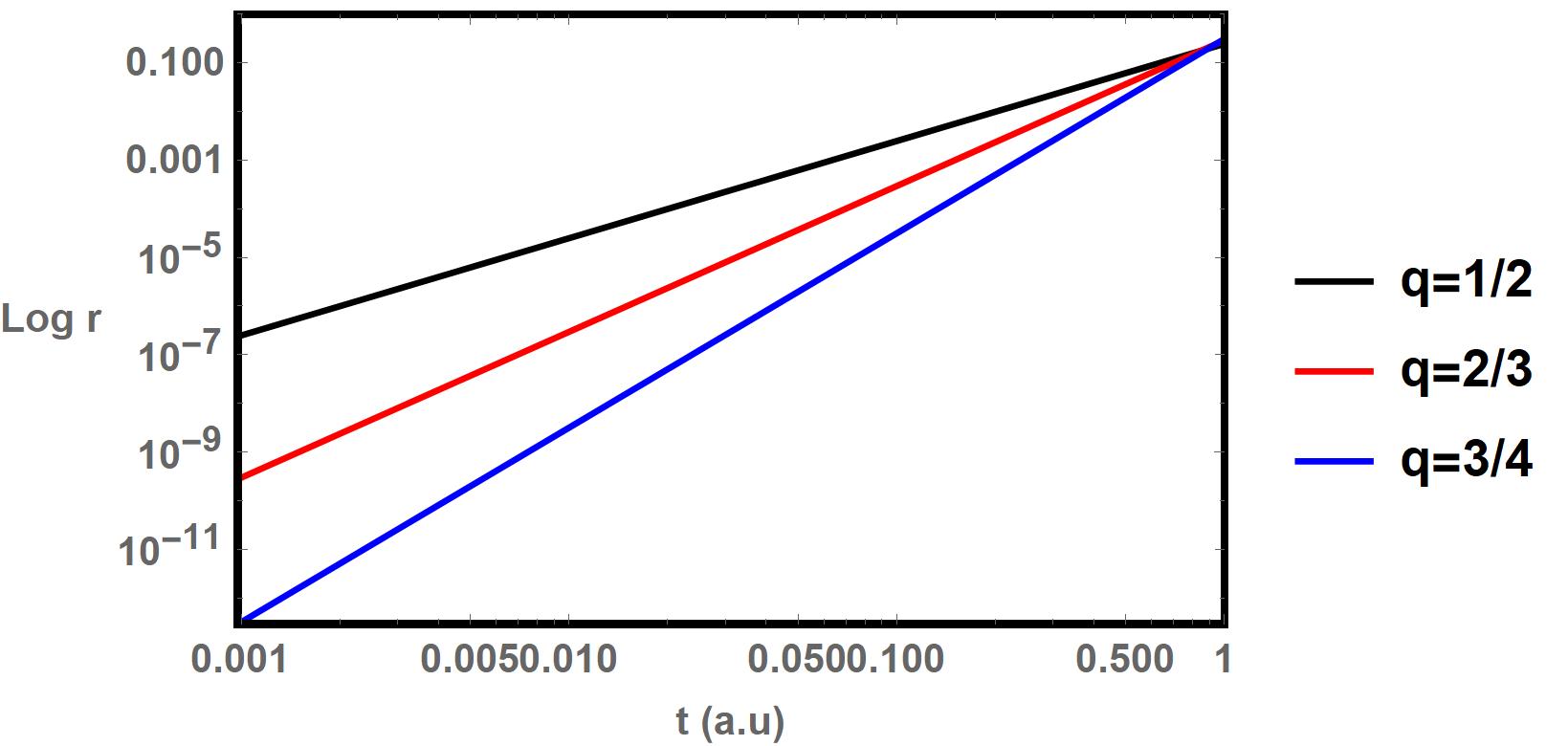}
	\caption{ Log-plot for different $q$ values  Eq. \eqref{PowerLaw}  }
	\label{Fig:figura2}
\end{figure}
\subsection{Entropic power law  growth}
Using the same previous dimensionless definitions Eq. \eqref{universal} is a follows   
\begin{align}
r^{1-q} = 1 - (1-r_{0}^{1-q})e^{-q\tau},
\end{align}
for early stages $e^{-q\tau} \approx 1 - q\tau$ and $r_{0} \rightarrow 0$, we obtain  a power growth law  behavior
\begin{align}
\label{PowerLaw}
r \approx c t ^D,
\end{align}
where $c = (qk)^D$ and 
\begin{equation}
\label{qexponent}
D \equiv  \frac{1}{1-q}.
\end{equation}  
The relationship between mass and time with statistics for growth in the early stages can be observed and can be related to a particular  exponent growth.
\\
In tumor growth, this kind of power law has been reported experimentally   \cite{Bru} and verified by numerical simulations \cite{Menchon}.\\
This kind of  growth law  has been discussed to explain the empirical evidence of tumor growth and the process of avascular tumors \cite{Renato}.\\
In order to illustrate, let us consider some examples, substituting into \eqref{PowerLaw} some $q$ values. For $q = \frac{1}{2}$ it results in a parabolic growth  $r \propto   t ^2$,  a cubic growth $r \propto  t ^3$ can be obtained using $q = \frac{2}{3}$, and so on.\\
For instance, the exponent $3/4$ is known to be associated to space-filling fractal patterns, while an exponent $2/3$ is associated to diffusion controlled growth and is observed for birds and small mammals \cite{Dodds,Craig}.\\
Some authors argued, that the exponent 2/3 may control tumor growth in its proangiogenic phase, and $3/4$ is likely to be more adequate to full-blown vascularized tumor expansion \cite{Condat1}.
The connection with the Bertalanffy-type model  ($2/3$ exponent) their  features and mechanics in  metabolism energy interchange was reported\cite{Roeder}. 
\\
The power law description with  non integer $q$ is related to the optimization of energy transport thought  thermodynamic  analysis leads to the evolution of fractal-like distribution networks.\\
Recently, using experimental data  collected  for mammals, birds, insects and plants several power laws  has been reported in  \cite{Fernando}.
\\
Here, it is important to note that  Eq. \eqref{qexponent} the exponent can not be mathematically defined for $q = 1$.  In Figure ~ \ref{Fig:figura2}, we  show that the power law behavior of Eq. \eqref{PowerLaw} for different $q$ values.
It is worth emphasizing that all pure power laws would imply an unbounded growth, which is impossible from the biophysical point of view.\\
However, this does not deny its validity and utility for a power laws growth model. It can provide information for early stages of tumors and can be applied successfully in radiotherapy treatments \cite{Sotolongo1,Sotolongo2}.

\section{Conclusions}
We presented a new  entropic growth model in terms of non extensive statistics. This
model has the mathematical shape of the “universal growth law” in the sense of West \cite{West}.
We have shown that for different $q$ values, the classical growth models can
be obtained now in terms of nonadditive entropy . We found that statistic growth
laws approximately exhibit a power-law behavior for a general class of phenomena.
This work can also motivate the use of nonextensive thermodynamics in biophysics
molecular modeling  \cite{Umberto}, where a detailed study was done using cellular heat exchange during 
metastasis. This model can be used to predict the growth of the COVID-19, where
the parameter $q$ can be   related to the nature and structure of the virus.

\section*{Acknowledgement}
IR  and JL was supported by  the Basic Science Research Program
through the National Research Foundation of Korea (NRF) funded by the Ministry of 
Science and ICT [NRF-2017R1E1A1A01077717].

\end{document}